\documentclass[conference]{IEEEtran}
\usepackage{cite}
\usepackage{amsmath,amssymb,amsfonts}
\usepackage{acronym}
\usepackage{algorithmic}
\usepackage{bbm}
\usepackage{bbold}
\usepackage{acronym}
\usepackage{graphicx}
\usepackage{physics}
\usepackage{textcomp}
\usepackage{xcolor}
\usepackage{subfig}

\let\oldthebibliography\thebibliography
\renewcommand{\thebibliography}[1]{\oldthebibliography{#1}\setlength{\itemsep}{-0.28em}}

\def\BibTeX{{\rm B\kern-.05em{\sc i\kern-.025em b}\kern-.08em
    T\kern-.1667em\lower.7ex\hbox{E}\kern-.125emX}}

\def\BibTeX{{\rm B\kern-.05em{\sc i\kern-.025em b}\kern-.08em
    T\kern-.1667em\lower.7ex\hbox{E}\kern-.125emX}}
\begin{document}

\title{Identifiability in Quantum State, Process, and Network Tomography 
\thanks{Identify applicable funding agency here. If none, delete this.}
}

\author{\IEEEauthorblockN{Athira Kalavampara Raghunadhan\textsuperscript{1},  Matheus Guedes De Andrade\textsuperscript{2}, Don Towsley\textsuperscript{2}, Indrakshi Dey\textsuperscript{3},\\   Daniel Kilper\textsuperscript{1}, Nicola Marchetti\textsuperscript{1} }
\\
\IEEEauthorblockA{\textsuperscript{1}CONNECT Research Centre, School of Engineering, Trinity College Dublin, Ireland}
\IEEEauthorblockA{\textsuperscript{2}Manning College of Information and Computer Science, University of Massachusetts Amherst, USA}
\IEEEauthorblockA{\textsuperscript{3}Department of Computing, School of Science, South East Technological University, Waterford, Ireland}
}

\newacro{QNT}{Quantum Network Tomography}
\newacro{QST}{Quantum State Tomography}
\newacro{QPT}{Quantum Process Tomography}
\newacro{IC}{Informationally Complete}
\newacro{FIM}{Fisher Information Matrix}
\newacro{CRB}{Cramér-Rao Bound}

\maketitle

\section{Background and Objectives}
\begin{figure}[!t]
\centering
\includegraphics[scale=0.45]{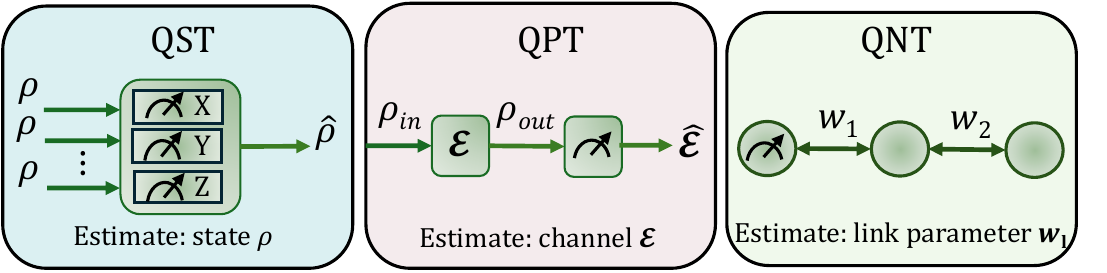}
\vspace{-0.5em}
\caption {Quantum State, Process, and Network Tomography}
\vspace{-0.5em}
\label{0}
\end{figure}

\ac{QST}, \ac{QPT}, and \ac{QNT} are related parameter-estimation problems that aim to reconstruct different physical quantities, as illustrated in Fig.~\ref{0}. \ac{QST} estimates an unknown quantum state, represented by its density matrix, from the measurement outcomes \cite{PhysRevA.64.052312}. QPT characterizes an unknown quantum channel using known input states and measurements of the corresponding outputs \cite{PhysRevA.77.032322}. \ac{QNT}, in contrast, aims to infer parameters associated with individual links from end-to-end probe measurements collected at accessible monitor nodes \cite{qntnetwork, optimalqnt}.

A key distinction among the three tomography problems lies in the conditions required to achieve identifiability, the ability to determine unknown parameters uniquely from the available measurement statistics. In \ac{QST} and \ac{QPT}, the experimenter can choose an \ac{IC} measurement set. \ac{QNT} limits the reachable measurements to what topology and monitor placement allow, so the admissible probe paths fix the information available about the link parameters. 

This work studies all three tomography problems through a common \ac{FIM} \cite{10.1145/2796314.2745862}. We factorize \ac{QNT} \ac{FIM} and show that its rank equals the rank of the path-link incidence matrix at every interior parameter value, so local and global identifiability coincide. We then show that \ac{QST} and \ac{QPT} attain full rank under \ac{IC} settings, \ac{QNT} loses rank when the probe paths leave link parameters indistinguishable, and increasing the number of copies scales the \ac{FIM} eigenvalues while leaving its rank fixed.

\section{Methodology}
Consider a quantum network represented by an undirected graph $G=(V,E)$, where $V$ is the set of quantum processors, and $E=\{e_1,\ldots,e_L\}$ is the set of quantum links. For the \ac{QNT} analysis, each link $e_l$ is modeled as a single-qubit depolarizing channel,
$\mathcal{D}_{w_l}(\rho) = w_l\rho+(1-w_l)\frac{I}{2}$, for $0<w_l<1$
where $w_l$ is the unknown link parameter. For \ac{QST} and QPT, we consider a general two-qubit state and a general single-qubit CPTP channel, respectively. To assess identifiability across \ac{QST}, QPT and \ac{QNT}, we use \ac{FIM}, which quantifies the measurement information about the unknown parameters. For a parameter vector $\boldsymbol{\theta}$ and measurement outcomes $y$ with probabilities $p_y(\boldsymbol{\theta})$, \ac{FIM} is
\begin{equation}
\label{fim}
[F(\boldsymbol{\theta})]_{ij}
=
\sum_y
\frac{1}{p_y(\boldsymbol{\theta})}
\frac{\partial p_y(\boldsymbol{\theta})}{\partial\theta_i}
\frac{\partial p_y(\boldsymbol{\theta})}{\partial\theta_j}.
\end{equation}
The \ac{FIM} bounds achievable precision through the \ac{CRB}, so full rank is the condition for a finite bound along every parameter direction, and rank deficiency exposes directions the statistics leave undetermined. 

For \ac{QST}, the general two-qubit state $\rho(\boldsymbol{\phi})$ is described by $15$ independent real parameters in the Pauli basis. For a POVM $\{E_m\}$ performed on the two-qubit state $\rho(\boldsymbol{\phi})$, the probability of the outcome $m$ is
$
p_m(\boldsymbol{\phi}) = \operatorname{Tr}\!\left[E_m\rho(\boldsymbol{\phi})\right].
$
Hence, the \ac{QST} \ac{FIM} is
\begin{equation}
[F_{\mathrm{QST}}]_{ij}
=
\sum_m
\frac{
\operatorname{Tr}\!\left[E_m\frac{\partial\rho}{\partial\phi_i}\right]
\operatorname{Tr}\!\left[E_m\frac{\partial\rho}{\partial\phi_j}\right]
}{
\operatorname{Tr}[E_m\rho]
}.
\end{equation}

For the \ac{QPT} comparison, the general single-qubit CPTP channel in the affine Bloch representation, $\mathbf{r}^{\mathrm{out}} = T \mathbf{r}^{\mathrm{in}} + \mathbf{t},$ where $T \in \mathbb{R}^{3 \times 3}$ and $\mathbf{t} \in \mathbb{R}^3$, giving 12 real channel parameters. For a known input state $s$ with the Bloch vector $\mathbf{r}_s = (r_{s,x}, r_{s,y}, r_{s,z})^{\top}$, define
$\mathbf{a}_s = (1, r_{s,x}, r_{s,y}, r_{s,z})^{\top}$, and $\boldsymbol{\theta}_k = (t_k, T_{kx}, T_{ky}, T_{kz})^{\top},$ for $k \in \{x, y, z\}$. Measurement of output in the Pauli-$k$ basis gives
$
p_{\pm|s,k} = \frac{1 \pm \mathbf{a}_s^{\top} \boldsymbol{\theta}_k}{2}.
$
Substituting these probabilities into Eq.~\eqref{fim}, the \ac{FIM} associated with the Pauli-$k$ output measurement is
\begin{equation}
F_k^{\mathrm{QPT}} = \sum_s \frac{\mathbf{a}_s \mathbf{a}_s^{\top}}{1 - (\mathbf{a}_s^{\top} \boldsymbol{\theta}_k)^2}.
\end{equation}
The three parameter blocks are disjoint, and each Pauli measurement depends on one block. Thus, $F_{\mathrm{QPT}} = \operatorname{diag}\!\left(F_x^{\mathrm{QPT}},F_y^{\mathrm{QPT}},F_z^{\mathrm{QPT}}\right)$, gives the total \ac{QPT} \ac{FIM}. Four input states whose $\mathbf{a}_s$ span $\mathbb{R}^4$, such as $\{|0\rangle, |1\rangle, |+\rangle, |{+}i\rangle\}$, with Pauli-$X$, $Y$, and $Z$ output measurements, make every block full rank and give an \ac{IC} \ac{QPT} configuration.

For \ac{QNT}, we employ the joint Bell-state measurement method in \cite{optimalqnt}. Let $\mathbf{w}=[w_1,\ldots,w_L]$ denote the unknown link parameters. For a probe path $P$, cyclic state generation produces a Bell-diagonal state with an effective Werner parameter
$
X_P = \prod_{e_l\in P} w_l^2,
$
since each link contributes twice to the cyclic probe. A Bell-basis measurement on this state gives
$
p_{\Phi^+}^{P} = \frac{1+3X_P}{4}$, and $p_{\Phi^-}^{P} = p_{\Psi^+}^{P} = p_{\Psi^-}^{P} = \frac{1-X_P}{4}.
$
For a set of $M$ probe paths, let $\mathbf{A}\in\{0,1\}^{M\times L}$ denote the path-link incidence matrix, with $A_{ml}=1$ when the path $m$ traverses link $e_l$, and $\mathbf{X}$ collect the corresponding $X_P$. Then
$
\log \mathbf{X} = 2\mathbf{A} \log \mathbf{w},
$
so, for $0<w_l<1$, the link parameters are uniquely identifiable if and only if $\operatorname{rank}(\mathbf{A})=L$. Applying Eq.~\eqref{fim} to these probabilities and differentiating through $X_P$, the contribution of path $P$ is
\begin{equation}
F_{ij}^{P} = \frac{3\,g_i^{P}g_j^{P}}{(1+3X_P)(1-X_P)},\quad
g_i^{P}=\frac{\partial X_P}{\partial w_i}=\frac{2A_{Pi}X_P}{w_i},
\end{equation}
which reduces to $12w_i^2/[(1+3w_i^2)(1-w_i^2)]$ for a directly monitored link $P=\{e_i\}$. Summing over the paths gives $F_{\mathrm{QNT}}=\mathbf{G}^{\top}\mathbf{C}\mathbf{G}$, with $\mathbf{G}=2\operatorname{diag}(X_P)\mathbf{A}\operatorname{diag}(w_l^{-1})$ and $\mathbf{C}=\operatorname{diag}(3/[(1+3X_P)(1-X_P)])$. For $0<w_l<1$, the diagonal matrices multiplying $\mathbf{A}$ are non-singular and $\mathbf{C}$ is positive definite. Hence, $\operatorname{rank}(F_{\mathrm{QNT}})=\operatorname{rank}(\mathbf{A})$, so the full-rank condition of \ac{FIM} agrees with the global identifiability condition of the log-linear model. All FIM expressions above correspond to one copy per measurement configuration, denoted by $F_1$. For $N$ independent copies per configuration, $F_N = N F_1$.

\section{Key Results}
The \ac{QST} and \ac{QPT} models are evaluated at $\rho=0.9|\Phi^+\rangle\langle\Phi^+|+0.1\,I_4/4$ and at $T=0.9I$, $\mathbf{t}=0$, while \ac{QNT} uses $\mathbf{w}=(0.8,0.9,0.85)$. Table~\ref{tab:identifiability} summarizes the findings. Nine local Pauli settings $P\otimes Q$ with $P,Q \in \{X,Y,Z\}$ form an \ac{IC} set for \ac{QST} and give a full-rank \ac{FIM} for the 15 state parameters ($\lambda_{\min}=1.00$); the \ac{IC} \ac{QPT} configuration gives a full-rank \ac{FIM} for the 12 channel parameters ($\lambda_{\min}=0.49$).
\begin{table}[htbp]
\centering
\caption{FIM-based identifiability comparison. $n_\theta$ is the number of unknown parameters.}
\label{tab:identifiability}
\footnotesize
\setlength{\tabcolsep}{4pt}
\begin{tabular}{|c|c|c|c|c|c|}
\hline
\textbf{METHOD} & $\boldsymbol{\theta}$ & $\boldsymbol{n_\theta}$ & $\boldsymbol{\operatorname{rank}(F)}$ &\textbf{IDENTIFIABLE} \\
\hline
QST (\ac{IC}) & $\phi$ & 15 & 15 & Yes \\
\hline
QPT (\ac{IC}) & $T, t$ & 12 & 12 & Yes \\
\hline
QNT ($P_{12},P_{13},P_1$) &  $w$ & 3 & 3 & Yes \\
\hline
QNT ($P_{12},P_{13}$) & $w$ & 3 & 2 & No \\
\hline
\end{tabular}
\end{table}
\vspace{-1.5mm}

\begin{figure}[t!]
\vspace{-3mm}
\centering
\hfill
\subfloat[\label{q22}]{\includegraphics[scale=0.34]{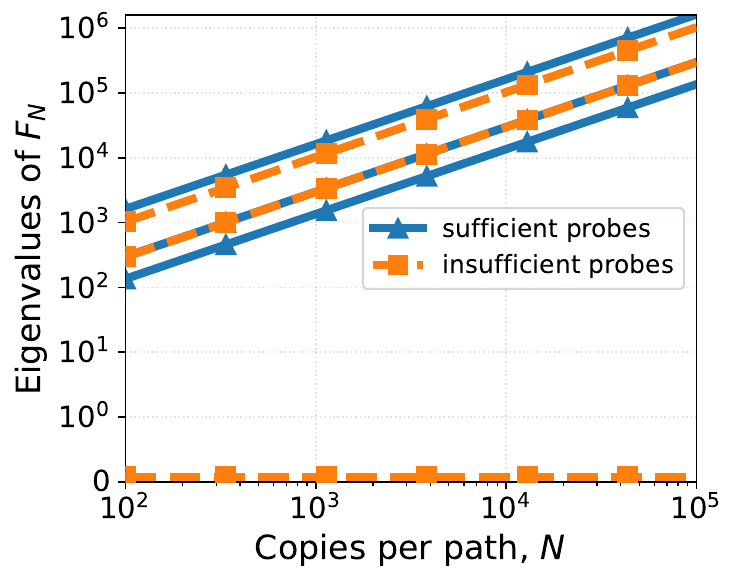}}%
\hfill
\subfloat[\label{q23}]{\includegraphics[scale=0.34]{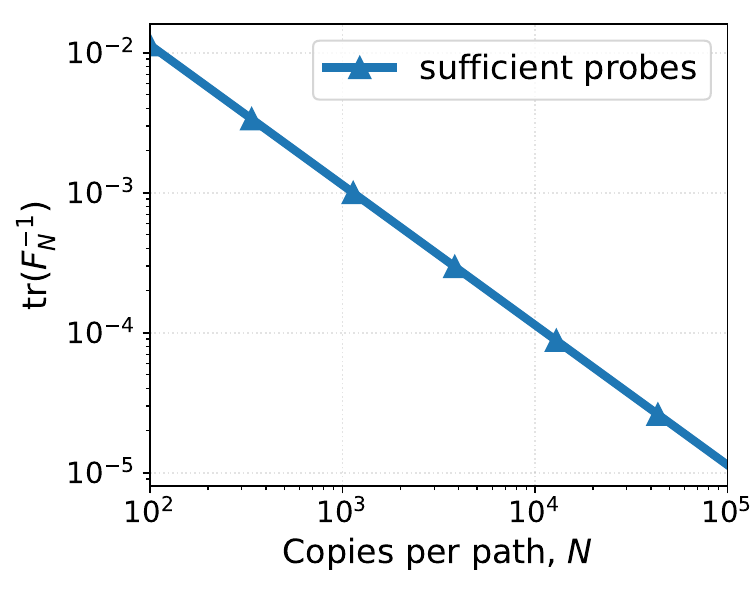}}%
\vspace{-0.5em}
\caption{(a) Eigenvalues of $F_N$ vs $N$ for both \ac{QNT} probe sets, (b) trace of $F_N^{-1}$ vs $N$ for the sufficient probe set.}
\label{2}
\end{figure}

For \ac{QNT} we take a three-link star network with a monitor placed at leaf node connected to $e_1$, probed by $P_{12}=\{e_1,e_2\}$ and $P_{13}=\{e_1,e_3\}$. They cover all three links and give $\mathbf{A}=\left[\begin{smallmatrix}1&1&0\\1&0&1\end{smallmatrix}\right]$ with $\operatorname{rank}(\mathbf{A})=2$ and \ac{FIM} eigenvalues $\{0,\,2.96,\,10.26\}$. The null direction $\delta\mathbf{w}\propto(w_1,-w_2,-w_3)$ corresponds to the rescaling $w_1\!\mapsto\!\alpha w_1$, $w_{2,3}\!\mapsto\!w_{2,3}/\alpha$, which leaves $X_{P_{12}}$ and $X_{P_{13}}$ unchanged and therefore leaves every outcome probability unchanged, leaving the link parameters unidentifiable for any number of copies. Adding the direct probe $P_1=\{e_1\}$ raises $\operatorname{rank}(\mathbf{A})$ to $3$ and lifts $\lambda_{\min}$ to $1.36$. Identifiability in \ac{QNT} therefore rests on the probe-path set on top of the measurement-completeness condition that \ac{QST} and \ac{QPT} impose.
Since $F_N=NF_1$, extra copies scale every eigenvalue by $N$ and leave the \ac{FIM} rank fixed. Fig.~\ref{q22} shows this: with two probe paths, one eigenvalue of $F_N$ remains zero for all $N$, so \ac{FIM} stays singular and no finite \ac{CRB} exists along the corresponding null direction. With three probe paths, every eigenvalue grows linearly in $N$ and $\operatorname{tr}(F_N^{-1})$ decays as $1/N$, as shown in Fig.~\ref{q23}, indicating improved estimation precision.
\vspace{-0.5em}
\section{Significance and Impact}
This work provides a \ac{FIM}-rank-based perspective on identifiability in \ac{QST}, \ac{QPT}, and \ac{QNT}, showing how the estimation problem changes when moving from state and channel characterization to network-level inference. \ac{QNT} adds a structural requirement on top of measurement completeness, since additional copies leave the \ac{FIM} rank unchanged. Reliable network characterization therefore requires not only sufficient measurement resources, but also probe paths that provide independent information about the underlying links. These results provide a basis for assessing identifiability across quantum tomography problems and motivate further research in \ac{QNT}, where topology and monitor placement constrain end-to-end probe design, which is essential for link inference without direct access to every network link.

\textbf{Keywords}---Quantum Tomography, Identifiability, Fisher Information Matrix. 
\vspace{-0.5em}
\section*{Acknowledgments}
This research was supported by Research Ireland grant 21/US-C2C/3750 for CoQREATE, CONNECT-2 grant 13/RC/2077\_P2, and by the NSF-ERC Center for Quantum Networks grant EEC- 1941583.
\vspace{-0.5em}
\bibliographystyle{unsrt}
\bibliography{references}

@INPROCEEDINGS{optimalqnt,
  author={Raghunadhan, A. K. and De Andrade, M. G. and Towsley, D. and Dey, I. and Kilper, D. and Marchetti, N.},
  booktitle={ICC 2026 - IEEE International Conference on Communications}, 
  title={Measurement Strategies and Estimation Precision in Quantum Network Tomography}, 
  year={2026},
  volume={},
  number={},
  pages={1-6},
  doi={10.1109/ICC59461.2026.11588067}}

@article{10.1145/2796314.2745862,
author = {He, T. and Liu, C. and Swami, A. and Towsley, D. and Salonidis, T. and Bejan, A. I. and Yu, P.},
title = {Fisher Information-based Experiment Design for Network Tomography},
year = {2015},
issue_date = {June 2015},
publisher = {Association for Computing Machinery},
address = {New York, NY, USA},
volume = {43},
number = {1},
issn = {0163-5999},
url = {https://doi.org/10.1145/2796314.2745862},
doi = {10.1145/2796314.2745862},
journal = {SIGMETRICS Perform. Eval. Rev.},
month = jun,
pages = {389–402},
numpages = {14}
}

@article{PhysRevA.64.052312,
  title = {Measurement of qubits},
  author = {James, D. F. V. and Kwiat, P. G. and Munro, W. J. and White, A. G.},
  journal = {Phys. Rev. A},
  volume = {64},
  issue = {5},
  pages = {052312},
  numpages = {15},
  year = {2001},
  month = {Oct},
  publisher = {American Physical Society},
  doi = {10.1103/PhysRevA.64.052312},
  url = {https://link.aps.org/doi/10.1103/PhysRevA.64.052312}
}

@article{PhysRevA.77.032322,
  title = {Quantum-process tomography: Resource analysis of different strategies},
  author = {Mohseni, M. and Rezakhani, A. T. and Lidar, D. A.},
  journal = {Phys. Rev. A},
  volume = {77},
  issue = {3},
  pages = {032322},
  numpages = {15},
  year = {2008},
  month = {Mar},
  publisher = {American Physical Society},
  doi = {10.1103/PhysRevA.77.032322},
  url = {https://link.aps.org/doi/10.1103/PhysRevA.77.032322}
}

@article{qntnetwork,
  author  = {De Andrade, M. G. and Navas, J. and Guha, S. and Monta{\~n}o, I. and Raymer, M. and Smith, B. and Towsley, D.},
  title   = {Quantum Network Tomography},
  journal = {IEEE Network},
  volume  = {38},
  number  = {5},
  pages   = {114--122},
  year    = {2024},
  doi     = {10.1109/MNET.2024.3403805}
}

\end{document}